\documentclass[10pt,a4paper,aps,showpacs,amsmath,amssymb,prl,floatfix,twocolumn,reprint]{revtex4}

\usepackage{slashed}
\usepackage{xcolor}
\usepackage{hyperref}

\usepackage[pdftex]{graphicx}
\usepackage{subfigure}
\usepackage{grffile}
\usepackage{feynmp}

\DeclareSymbolFont{rsfs}{U}{rsfs}{m}{n}
\DeclareSymbolFontAlphabet{\mathrsfs}{rsfs}

\begin{document}

\title{Two-photon Compton process in pulsed intense laser fields}
\author{Daniel Seipt}
\email{d.seipt@hzdr.de}

\author{Burkhard K\"ampfer}
\email{kaempfer@hzdr.de}
\affiliation{Helmholtz-Zentrum Dresden-Rossendorf, POB 51 01 19, 01314 Dresden, Germany}

\pacs{12.20.Ds, 32.80.Wr, 41.60.-m}
\keywords{photon pairs, short intense laser pulses, Dirac-Volkov propagator}

\begin{abstract}
Based on strong-field QED in the Furry picture we use the Dirac-Volkov propagator
to derive a compact expression for the differential emission probability
of the two-photon Compton process in a pulsed intense laser field.
The relation of real and virtual intermediate
states is discussed, and the natural regularization of the on-shell contributions
due to the finite laser pulse is highlighted.
The inclusive two-photon spectrum is two orders of magnitude stronger than expected from a perturbative estimate.
\end{abstract}

\maketitle


\begin{figure}[!b]
\vspace*{-3mm}
\includegraphics[scale=1]{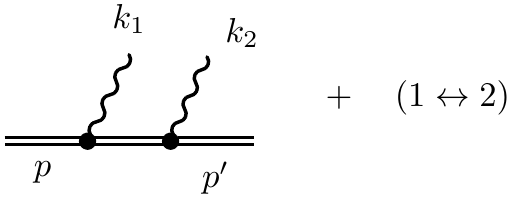}
 \caption{Feynman diagrams for the two-photon Compton process, where two photons (wavy lines) are emitted.
 The double lines representing laser dressed Volkov in and out states and the Dirac Volkov propagator between the two vertices, respectively. \vspace*{-5mm}}
 \label{fig.feynman}
\end{figure}

Recently, experiments on two-photon emission by electrons in an intense laser field have been proposed~\cite{Chen:Tajima:PRL1999,Thirolf:EPJD2009}.
The motivation of such experiments is seen in~\cite{Chen:Tajima:PRL1999,Schutzhold:PRL2008} in an attempt to {verify the Unruh radiation \cite{Unruh:PRD1984,Crispino}}
which is related to the physical vacuum experienced by accelerated observers
in a flat space-time and manifests as the emission of entangled photon pairs off accelerated charges.
The relation between the QED two-photon Compton process
and {Unruh radiation} has been analyzed further in \cite{Schutzhold:EPJD2009}.
As the intensity of the Unruh radiation increases with the acceleration \cite{Schutzhold:PRL2006}, e.g.~caused by a strong electromagnetic field, one is naturally lead to
think on strong laser fields as driving force.
High field strengths are nowadays achieved by the temporal compression of laser pulses to a few femtoseconds.
The temporal structure of
the laser pulse has a strong impact on the nonlinear Compton spectrum \cite{Seipt:PRA2011}
(hereafter termed one-photon Compton scattering) and its cross channels~\cite{Heinzl:PLB2010,Ilderton:SinglePhoton:2011}.
%

The theoretical description of the two-photon Compton process
(also termed double Compton scattering) in the perturbative (weak-field) regime had been accomplished first in~\cite{Heitler:Physica1934,Mandl:Skyrme:1952}
and was verified experimentally soon afterwards
\cite{Bracci:NuovoCimento1956}; for a more
recent experiment
cf.~\cite{Sandhu:2000}.
%

In second-order strong-field processes~(e.g.~\cite{Jentschura:PRL,Loetstedt:2007,Krajewska:PRA2010,Hu:2010}), 
intermediate particles can become real (i.e.~go on their mass shell) due to the presence of the background field.
The on-shell contributions
have been discussed
\cite{Oleinik1,Roshchupkin:Review}
as
Oleinik resonance singularities.
The relation of the on- and off-shell processes
in the case of a photon propagator has been analyzed recently in~\cite{Hu:2010,Ilderton:2010}
for the trident pair production with respect to the interpretation of the SLAC E-144 data~\cite{Burke:PRL1997}.

In this paper, we provide a complete description of the nonperturbative two-photon Compton process in a
pulsed intense laser field. 
We show the
significant modification of the two-photon Compton process by
short laser pulses when compared to infinite plane-wave fields considered previously
\cite{Jentschura:PRL}.
For the first time, an exact description of this process involving
a temporally shaped laser-dressed Dirac-Volkov propagator
within the Furry picture is given.
We calculate the photon pair emission probability and compare it to the one-photon Compton probability,
finding a substantially increased two-photon yield in short intense laser pulses as compared
to perturbative estimates.

%
The Feynman diagrams in Fig.~\ref{fig.feynman} correspond to the $S$ matrix
\begin{align}
 S &= -ie^2 \int d^4xd^4y \overline \Psi_{p'}(y) 
 \slashed \epsilon_{2}^*
 	\frac{e^{ik_2 \cdot y}}{\sqrt{ 2\omega_2}}
		  \mathcal G(y,x)
		  \slashed \epsilon_1^* 
		  	\frac{e^{ik_1 \cdot x}}{\sqrt{2\omega_1 }}
		\Psi_p(x) \nonumber \\
		&
		  \quad + (1\leftrightarrow 2),
\label{eq.def.S.matrix}
\end{align}
where $(1\leftrightarrow 2)$ means exchange of photons $1$ and $2$ accounting for
the symmetrization of the two-photon wave function.
$\Psi_p$ ($\overline{\Psi}_{p'}$) is the Volkov wave function for an electron in the entrance (exit) channel with momentum $p$ ($p'$), and $\mathcal G(y,x)$ denotes the dressed electron propagator
\cite{Ritus:JSLR1985},
$k_i$ ($\epsilon_i$) stand for the four-momenta (four-polarizations) of the emitted photons.
A dot, e.g.~in $k_1 \cdot x$, denotes the scalar product of four-vectors, and
$\slashed \epsilon_i = \gamma \cdot \epsilon_i$.

We describe a linearly polarized laser pulse by the four-potential
$A_\mu = E_0 \epsilon_\mu  a(\phi) / \omega$ with polarization four-vector $\epsilon^\mu=(0,1,0,0)$
and four-momentum $k^\mu=(\omega,0,0,-\omega)$ with
$\epsilon\cdot k=0$, phase $\phi=k\cdot x$, $a(\phi) = g(\phi) \cos \phi$ and temporal envelope function $g(\phi)$ which vanishes as $|\phi|\to \infty$.
The
nonlinearity parameter is $a_0 = eE_0/m\omega$ with laser frequency $\omega$,
electron mass $m$, charge $e$, and peak electric field $E_0$.
The perturbative regime corresponds to $a_0 \ll 1$.

Both, $\Psi_p(x)$ and 
$\mathcal G(x,y)$,~depend on the Volkov matrix functions
$E_p(x) = \Gamma_p(\phi) \exp \{-i p\cdot x - if_p(\phi)\} $ with
$ {\Gamma_p =  1 + {e \slashed k \slashed A}/{(2k\cdot p)}} $ and the nonlinear phase
\begin{eqnarray}
f_p(\phi)
 &=& \alpha_p \intop_0^\phi d \phi'     a(\phi')  +  \beta_p \intop_0^\phi d \phi' a^2(\phi') 
\end{eqnarray}
with coefficients $\alpha_p = ma_0 (p\cdot \epsilon)/(p\cdot k)$ and $\beta_p = m^2 a_0^2 / (2 p\cdot k) $. We employ light cone coordinates with $p_\pm = p^0 \pm p^3$,
$\mathbf p_\perp = (p^1,p^2)$
and $\mathbf p = (p_+,\mathbf p_\perp)$ such that $k_-$ is the only non-vanishing
component of the laser four-momentum with $\phi = k_-x_+/2$ and
$d^4x = k_-^{-1} d\phi d x_- d\mathbf x_\perp$.


The integrations over the transverse and $-$ components provide
momentum conserving delta distributions, such that the S matrix \eqref{eq.def.S.matrix}
becomes
\begin{align}
  S 
    = &
   \frac{-i\pi^2 e^2     \delta^3(\mathbf p' + \mathbf k_1 + \mathbf k_2 - \mathbf p  )}{\sqrt{\omega_1 \omega_2 (k\cdot p) (k\cdot p')}}     
   \int  d\ell d\phi \, d\psi  \,  \nonumber \\
    \ &\times e^{i(s-\ell)\phi - i f_{P_1}(\phi) + i f_{p'}(\phi)}
     e^{i \ell\psi - i f_p(\psi ) + i f_{P_1}(\psi)} \nonumber \\
  \ & \times \bar u_{p'} \overline{\Gamma}_{p'}(\phi) \slashed \epsilon_2^*
   \Gamma_{P_1}(\phi) 
    \frac{\slashed P_1 + \ell \slashed k +m}{(P_1 +\ell k)^2 - m^2 + i \varepsilon} \nonumber \\
  \ & \times \overline{\Gamma}_{P_1}(\psi) \slashed \epsilon_1^*
    \Gamma_p(\psi) u_p   
  + \ (1\leftrightarrow 2) , 
    \label{eq.S.integral.n}
\end{align}
with $P_1 = p-k_1$ and $s = (p'_-+k_{1-} + k_{2-} - p_-)/k_- > 0$.
For certain pulse profiles a part of the non-linear phases $f_p$ referring to the ponderomotive energy might be rewritten to generate explicitly a mass contribution in the denominator, leading to the familiar form of the propagator with a mass-shift $\Delta m^2 = m^2 a_0^2 / 2$ in infinite plane waves, e.g.~in \cite{Jentschura:PRL}. For our purposes, (\ref{eq.S.integral.n}) is more suitable.
The variables $s$ and $\ell$ are continuous analogues to the number of exchanged photons.
The integral over $\ell$ in (\ref{eq.S.integral.n})
accumulates all possible paths the system may take.
It can be evaluated analytically by
the contour integration technique: The integrand has a single pole at $\ell=\ell_1 - i\varepsilon'$
with $\ell_1 = (m^2-P_1^2)/2k\cdot P_1$ and
$\varepsilon' = \varepsilon \, {\rm sign} ( P_{1+})$.
To safely apply the residue theorem, one has to transform the integrand according to $\ell/(\ell-\ell_1+i\varepsilon') \to 1 + \ell_1/(\ell-\ell_1+i\varepsilon')$
such that the nontrivial part goes to zero as $\ell \to \infty$.
The result of this procedure,
\begin{align}
& \int_{-\infty}^\infty \! \! d\ell \frac{(\slashed P_1 + \ell \slashed k +m) e^{-i\ell (\phi-\psi)} }{\ell-\ell_1 + i\varepsilon'} = 
 2\pi \delta(\phi-\psi)\slashed k  \nonumber \\
& \ \quad  -2\pi i \theta(\phi-\psi ) e^{-i(\ell_1-i\varepsilon')(\phi-\psi)} (\slashed P_1 + \ell_1\slashed k + m), \
\label{eq.n.integration}
\end{align}
takes the structure of a fermion propagator in light-front form \cite{Kogut:1970} due to
the integration over the light-cone component $k_-$.
In the second line~of (\ref{eq.n.integration})
a time ordering (in the laser phase) is introduced
in the sense that the emission at
the second vertex has to happen at a later time than the emission at the first vertex
by means of the step function $\theta(\phi-\psi)$.
The quantity $\ell_1$ controls the amount of momentum $\ell_1 k$ transferred to the electron such that 
the propagator momentum is on its mass shell, i.e.~$(P_1 +\ell_1 k)^2=m^2$.
Additionally (\ref{eq.n.integration}) includes a part $\propto \delta(\phi-\psi)$ where both photons are emitted simultaneously.
This "instantaneous propagator", which is also known as light-front zero-mode propagator,
is specific to the fermion propagator and does not appear in the analysis of the trident process \cite{Ilderton:2010} with the photon propagator.
The propagator pole in (\ref{eq.n.integration}) always lies below the real axis due to momentum conservation. In particular, we have $p_+' = P_{1+} - k_{2+} > 0$ with $k_{2+} > 0$ and therefore $P_{1+}>0$.
Negative values of $P_+$ would shift the pole to the
upper half plane corresponding to the opposite time ordering $\theta(\psi-\phi)$.

The six-fold differential probability of two-photon emission per incident laser pulse finally reads
\begin{align}
d^6W &= \frac{\alpha^2
\big|
\mathrsfs M + (1\leftrightarrow 2)
\big|^2
}{64 \pi^4 (k\cdot p) (k\cdot p')}
\prod_{i=1}^2 \omega_i d\omega_i d\Omega_i,
\label{eq.number.pairs} \\
\mathrsfs M &= \frac{1}{2 k\cdot P_1} 
\left\{
 \sum_{n=0}^2 A_n(s) \bar u_{p'}  T_n u_p 
 \right. \nonumber \\
& 
 \left. \qquad \qquad
 - i \sum_{n,l=0}^2 B_{nl}(s,\ell_1)  \bar u_{p'}  U_{nl} u_p 
\right\}  , \label{eq.def.M}
\end{align}
with $\alpha= e^2/4\pi$ and
the phase integrals
\begin{align}
A_n (s) &= \int d\phi \, a^n(\phi) e^{is\phi -i f_p(\phi) + i f_{p'}(\phi)}, \label{eq.def.A}\\
B_{nl}(s,\ell_1) &= \int d\phi \, d\psi \, \theta(\phi-\psi) a^n(\phi) a^l(\psi) \label{eq.def.B}   \\
& \times  e^{i(s-\ell_1)\phi -if_{P_1}(\phi) + i f_{p'}(\phi)}
e^{i\ell_1 \psi - i f_p(\psi) + i f_{P_1}(\psi)},\nonumber
\end{align}
and the Dirac structures
$T_0 = \slashed \epsilon_2^* \slashed k \slashed \epsilon_1^*$,
$T_1 =  \overline X_{p'} \slashed \epsilon_2^* \slashed k \slashed \epsilon_1^* 
+   \slashed \epsilon_2^* \slashed k \slashed \epsilon_1^* X_p
 $,
$T_2 =  4 d_p d_{p'} (\epsilon_2^* \cdot k) (\epsilon_1^*\cdot k)  \slashed k $, 
$U_{00} = \slashed \epsilon_2^* G_1 \slashed \epsilon_1^* $,
$U_{01} =  \slashed \epsilon_2^* G_1 
(\overline X_{P_1}\slashed \epsilon_1^* + \slashed \epsilon_1^* X_p )$,
$U_{10} =  (\overline X_{p'}\slashed \epsilon_2^* + \slashed \epsilon_2^* X_{P_1}) G_1 
\slashed \epsilon_1^* $,
$U_{11} = (\overline X_{p'}\slashed \epsilon_2^* + \slashed \epsilon_2^* X_{P_1}) G_1 
(\overline X_{P_1}\slashed \epsilon_1^* + \slashed \epsilon_1^* X_p  )$,
$U_{02} = \slashed \epsilon_2^* G_1 \overline X_{P_1} \slashed \epsilon_1^* X_p$,
$U_{20} =  \overline X_{p'} \slashed \epsilon_2^* X_{P_1} G_1  \slashed \epsilon_1^* $,
$U_{12} =   (\overline X_{p'}\slashed \epsilon_2^* + \slashed \epsilon_2^* X_{P_1}) G_1 
\overline X_{P_1}\slashed \epsilon_1^* X_p$,
$U_{21} = \overline X_{p'}\slashed \epsilon_2^*  X_{P_1} G_1 
(\overline X_{P_1}\slashed \epsilon_1^* + \slashed \epsilon_1^* X_p  )$,
$U_{22} =  8 d_p d_{p'} d_{P_1}^2 (\epsilon_2^* \cdot k) (\epsilon_1^*\cdot k) (P_1\cdot k) 
 \slashed k $,
where the abbreviations $X_p = d_p \slashed k \slashed \epsilon$, $d_p = a_0 m / (2k\cdot p)$ and $G_1= \slashed P_1 + \ell_1 \slashed k + m$ are employed.
%
%
The phase integrals $A_0, B_{0l}, B_{n0}$ and $B_{00}$ are numerically non-convergent because of the missing pre-exponential factor.
However, these integrals can be defined as a superposition of convergent phase integrals by
applying a quantum gauge transformation
$\epsilon_i \to \epsilon_i + \lambda_i k_i$ ~\cite{Ilderton:2010},
yielding, e.g.
$
(s-\ell_1)B_{0l}(s,\ell_1) = 
 iA_l(s) + (\alpha_{P_1} - \alpha_{p'})B_{1l} (s,\ell_1)
+ (\beta_{P_1} - \beta_{p'}) B_{2l}(s,\ell_1).$
Thus, gauge invariance reduces the number of independent phase integrals from
twelve to six well-behaved ones.

Applying the Sokhotsky-Weierstrass theorem,
$(\ell-\ell_1 +i\varepsilon)^{-1} = \mathcal P (\ell-\ell_1)^{-1} - i \pi \delta(\ell-\ell_1)$, in (\ref{eq.n.integration})
one identifies the imaginary part of $\mathrsfs M$ as the contribution of real (on-shell) electrons with the aid of the delta distribution.
By recalling that principal value integration ($\mathcal P$) effectively means cutting out a small interval around zero it becomes clear that 
the real part of $\mathrsfs M$ refers to the virtual (off-shell) process. The phase integrals $A_n$
contribute only to the off-shell process.
Employing
the completeness relation for spinors
$G_1= \sum_\sigma u_{P_1+\ell_1k ,\sigma} \bar u_{P_1+\ell_1k,\sigma}$
to the numerator of the propagator 
($\sigma$ is the spin of the intermediate electron), the on-shell part
of $S$ yields a factorization on the amplitude level
\begin{align}
 S_{\rm on}  
 	=& \int \frac{d q_+ d^2\mathbf q_\perp}{2(2\pi)^3}
 	\sum_{\sigma = 1,2} 
 	S^{(1)}(q + (s_0-\ell_1) k \to p' + k_2)  \nonumber \\
 	& \times S^{(1)}(p + \ell_1 k \to q + k_1) 
 	+ (1 \leftrightarrow 2),
 \end{align}
where $S^{(1)}$ denotes the $S$ matrix for the one-photon Compton process with the given
energy-momentum conservation
(cf.~equation~(29) in \cite{Seipt:PRA2011}) integrated over all intermediate states.
Numerically, the interference between the two Feynman diagrams in Fig.~\ref{fig.feynman} turns out to be of the same order of magnitude as the non-interference terms.

In the weak-field limit $a_0 \ll 1$, the off-shell part of the $S$ matrix is proportional
to $a_0$ in lowest order corresponding to the absorption of one laser photon reproducing the known perturbative result.
Contrarily, the leading order contribution to the on-shell part is
$\propto a_0^2$, since at least two laser photons are needed to satisfy the on-shell energy momentum conservation.
In the limit of infinite laser waves
the on-shell part of the two-photon rate
gives rise
to Oleinik resonances \cite{Oleinik1}
at frequencies
\begin{eqnarray}
\omega_i^{\rm res}(\ell)	&=& \frac{\ell k\cdot p}{
\left[
	p + \Big(\ell + \frac{m^2a_0^2}{4k\cdot p} \Big) k
\right]
\cdot n_i},
\label{eq.resonances}
\end{eqnarray}
where $n_i=(1,\cos\varphi_i \sin \theta_i,\sin\varphi_i \sin\theta_i,\cos \theta_i)$
is the unit vector in the direction of $k_i$ (with
$\theta_i$ and $\varphi_i$ denoting the usual polar and azimuthal angles
of the emitted photons and integer $\ell$).
The resonances for $\omega_1$ ($\omega_2$) emerge when the propagator of the
first (second) Feynman diagram in Fig.~\ref{fig.feynman} comes on its mass shell.
A detailed discussion of the resonance
behavior will be given elsewhere; it has its own
right but does not affect the following results.
%

%
For our numerical evaluations we consider electrons with a Lorentz factor $\gamma = p^0/m = 10^4$, available, e.g.~at the European XFEL electron beam \cite{XFEL},
in head-on
collisions with the laser pulse.
Calculations have been performed for $a_0=1$ and a pulse
shape $g (\phi) = \cos^2 (\frac{\pi \phi}{2\tau})$ for $-\tau \geq \phi \geq \tau$ and zero otherwise, such that $\tau$ is the dimensionless FWHM pulse length with $\tau = 20$ corresponding to $9\rm \ fs$ FWHM for $\omega = 1.55 \ \rm eV$. (A similar kinematic situation with the same center-of-mass energy could be achieved by colliding an XFEL X-ray pulse \cite{XFEL} with low energy electrons, e.g. $\gamma = 10$, provided by an optical laser acceleration set-up \cite{Laseracc}.)
\begin{figure}[!ht]
\begin{center}
\hspace*{-2mm}
$
\vcenter{ \hbox{ \includegraphics[height = 4.1cm]{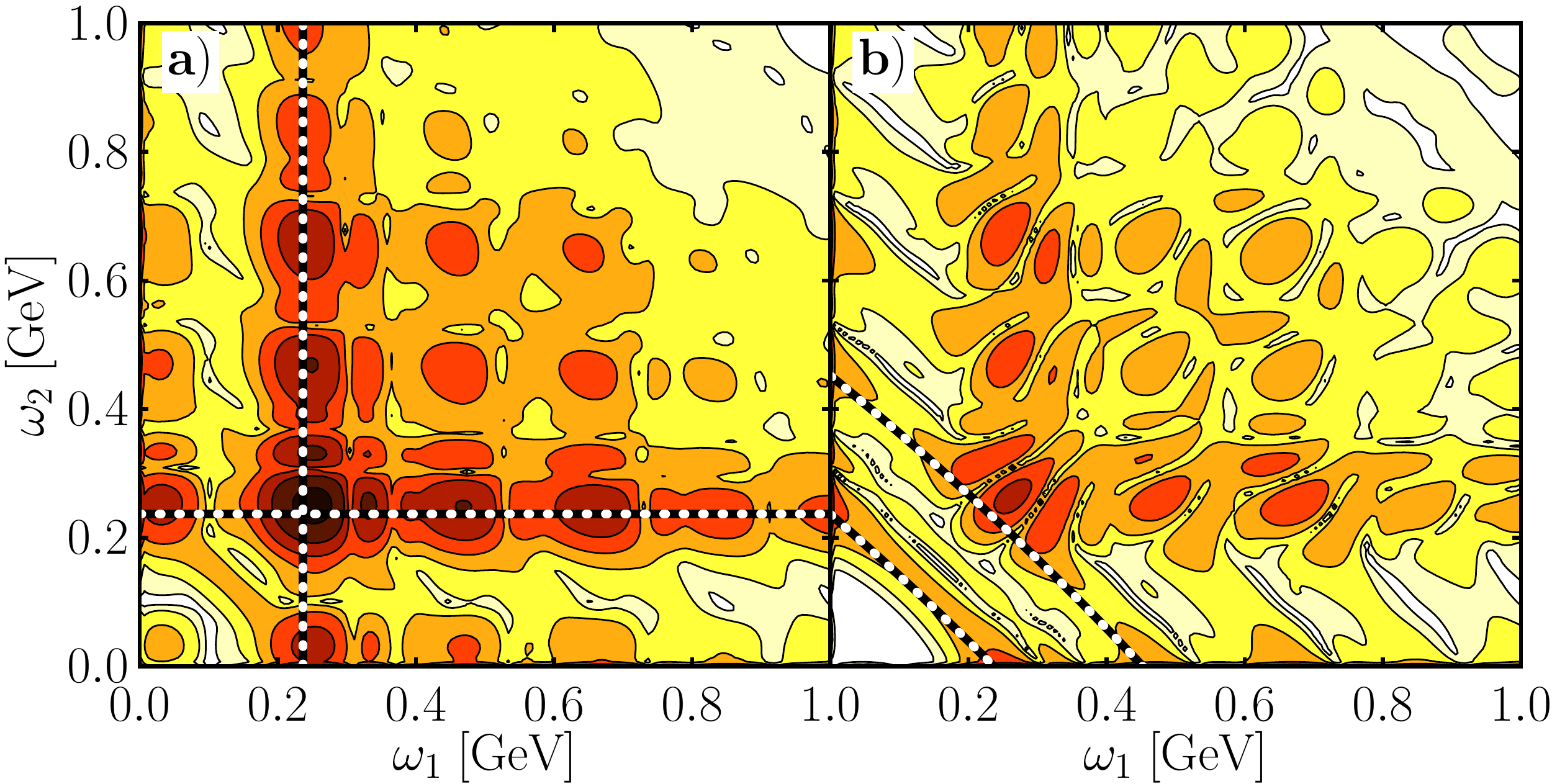} } }
\hspace*{-2.7mm}
\vcenter{ \hbox{ \includegraphics[height = 3cm]{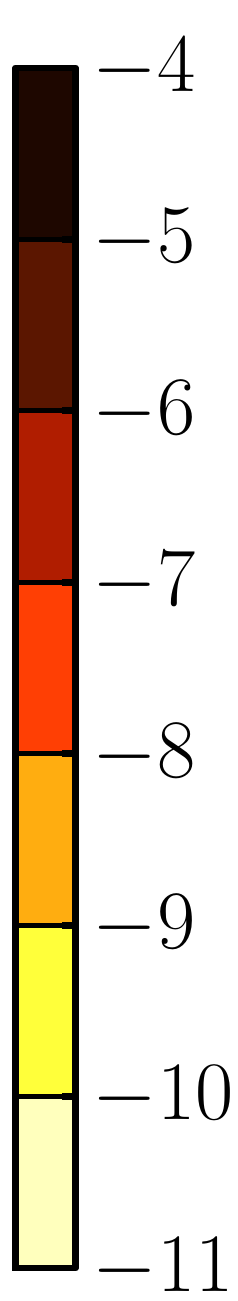}      \vspace{.5cm} } }
$
\end{center}
\vspace*{-6mm}
\caption{
Intensity distribution for two-photon emission in the
$\omega_1$-$\omega_2$ plane. We show the complete emission probability [$\bf a)$] and the off-shell contribution [$\bf b)$].
The color code represents the logarithm of the six-fold differential probability in eV,
$\log_{10} (d^6W/d\omega_1d\Omega_1 d\omega_2d\Omega_2 \rm \ [eV]) $.
For parameters see the text. \vspace*{-5mm}
}
\label{fig.omega_omega}
\end{figure}
In Fig.~\ref{fig.omega_omega}, results are exhibited for the differential
probability for two-photon emission as a function of $\omega_1$ and $\omega_2$.
Since the motion of the electron is relativistic, the radiation is produced in a cone
around the spatial direction of $p$ with a typical opening angle of $1/\gamma$. We show the
differential probability at angles
$\theta_{1,2} = 1/\gamma$ and $\varphi_1 = \pi/2$ and
$\varphi_2 = 3\pi/2$, i.e. the two photons are emitted in a plane perpendicular to the polarization plane of the laser.
The left panel in Fig.~\ref{fig.omega_omega} shows the complete differential probability as
the sum of on- and off-shell parts, while in the right panel the off-shell part is exhibited.
The probability distributions display complex characteristic patterns.
The differential spectrum in Fig.~\ref{fig.omega_omega} $\bf a)$ is dominated by the on-shell part in almost the whole $\omega_1$-$\omega_2$ phase space
for both $\omega_i > 200 \ \rm MeV$, where it is roughly one order of magnitude larger than the off-shell part. This is a generic feature also for different scattering angles.
The on-shell part shows a rectangular pattern, which is aligned parallel to the coordinate axes.
The spectrum has maxima in regions where the Oleinik resonances (\ref{eq.resonances}) would occur for infinite plane-waves and is particularly strong where both types of resonances intersect.
The resonances (\ref{eq.resonances}) for $\ell = 1$ are indicated in Fig.~\ref{fig.omega_omega}
$\bf a)$ as dotted lines.
The off-shell part exceeds the on-shell part for at least one of the $\omega_i$ below $200$ MeV, where
the maxima of the distribution are roughly aligned with the different harmonics $\ell$ of the infinite plane-wave energy correlation which read
\begin{eqnarray}
 \omega_2(\ell) = 
 \frac{
 \ell k\cdot p - p \cdot k_1 - 
 \Big(\ell + \frac{m^2 a_0^2}{4k\cdot p}\Big)  k \cdot k_1
 }
 {
\left[
p + 
\Big(\ell + \frac{m^2a_0^2}{4k\cdot p} \Big) k - k_1 
\right]\cdot n_2
 }\,;
 \label{eq.omega2}
\end{eqnarray}
these are shown as dotted lines for $\ell=1,2$ in Fig.~\ref{fig.omega_omega} {\bf b}) .
For higher photon energies
the pattern is more irregular but still symmetric with respect to an exchange of $\omega_1$
and $\omega_2$.
Fig.~\ref{fig.omega_omega} evidences the striking differences to the infinite plane-wave case:
The strict $\omega_2(\omega_1)$ correlation of (\ref{eq.omega2}) gets completely lost.
Instead, resonance like structures with subpeaks appear
which are produced by the ponderomotive broadening mechanism,
resembling the ones observed in the one-photon
Compton process~\cite{Seipt:PRA2011}.

\begin{figure}[!t]
\begin{center}
\vspace*{-3mm}
\includegraphics[width = 8.6cm]{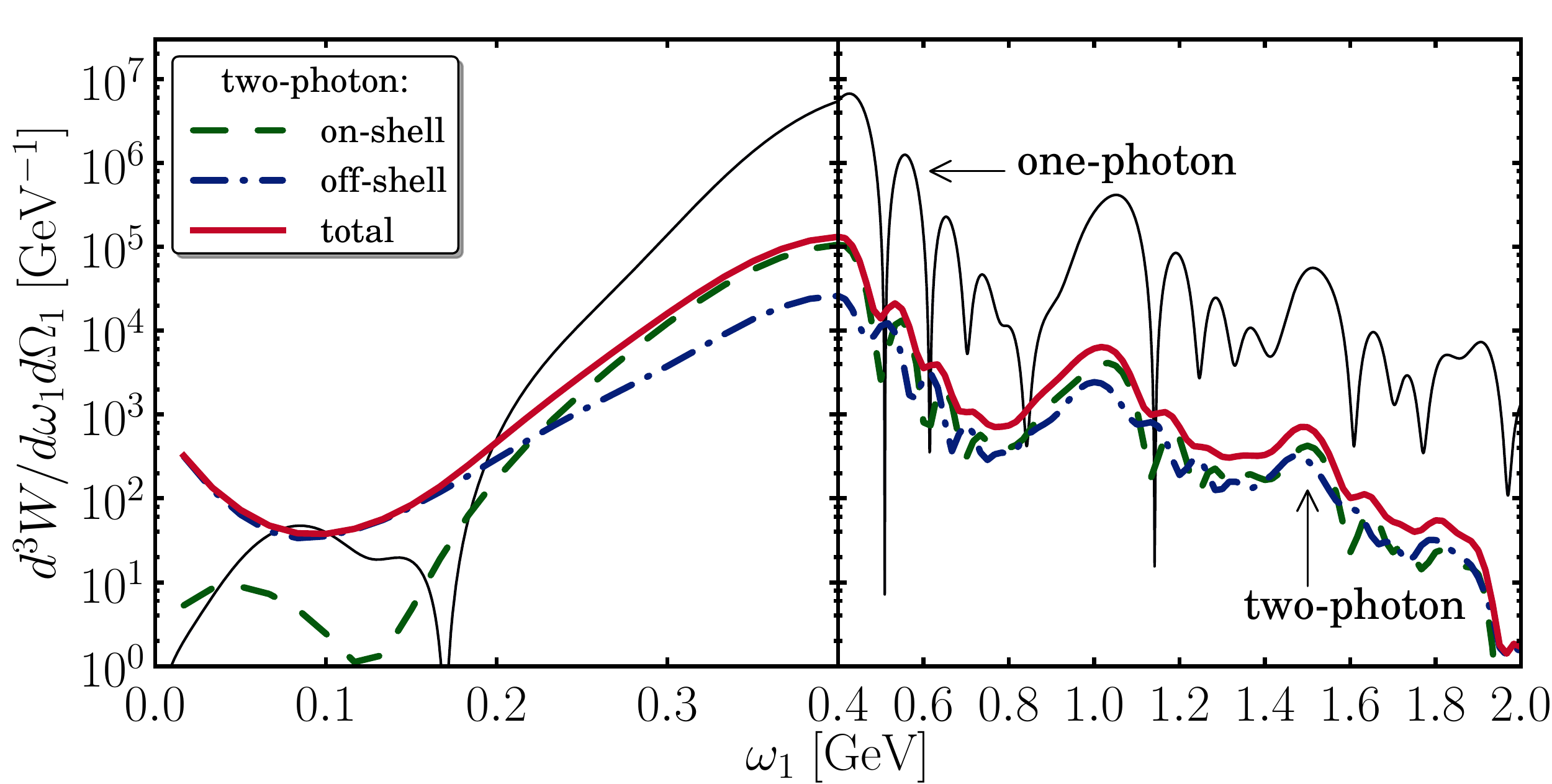}
\vspace*{-6mm}
\end{center}
\caption{Inclusive spectrum vs.~the one-photon Compton spectrum as a function of the
emitted photon frequency $\omega_1$. \vspace*{-5mm}}
\label{fig.inclusive_back}
\end{figure}

In Fig.~\ref{fig.inclusive_back} we exhibit the inclusive spectrum
$d^3W/d\omega_1 d\Omega_1$
arising from (\ref{eq.number.pairs}) by
integrating over the phase space of photon 2.
For soft photons $\omega_2\to 0$ the
emission probability of two-photon emission becomes divergent.
The cancellation of this infrared divergence by soft virtual photons
due to loop corrections of one-photon scattering is ensured by the Bloch-Nordsieck theorem \cite{Bloch:Nordsieck} as in the perturbative case.
For practical purposes, however, we include an infrared cutoff $\omega_2^{\rm min} = 100$ keV to avoid the soft-photon divergence in the spirit of~\cite{Jentschura:PRL}.
The value of the integral is rather insensitive to a variation of the cutoff in the range of $1-1000$ keV.

This inclusive spectrum accounts for the experimental observation of only one of the two photons.
To compare with one-photon Compton backscattering, we choose $\theta_1 = \varphi_1 = 0$.
In the case of strong laser fields,
e.g.~for $a_0=1$,
the inclusive spectrum
is found about two orders of magnitude below the one-photon spectrum for $\omega_1 > 200 \ \rm MeV$.
At photon energies $\omega_1<200$ MeV, the two-photon process exceeds the
one-photon process (see Fig.~\ref{fig.inclusive_back}), opening, at least in principle,
a window to access its observation without coincidence measurements.

{
Approximating
$\frac{dW}{d\omega_1} = \int \! d \Omega_1 \frac{dW}{d\omega_1 d\Omega_1}
\approx \frac{2\pi}{\gamma^2} \left.\frac{dW}{d\omega_1 d\Omega_1}
\right|_{
\genfrac{}{}{0pt}{3}{\theta_1=0}{\varphi_1=0}
}
$ and integrating over $\omega_1$,
one can estimate the total number of produced pairs as $1.1\times 10^{-3}$ per pulse and electron as compared to $5\times 10^{-2}$ coming from the single photon process.
With an assumed laser repetition rate of $10$ Hz
one can expect $950$ two-photon events
as compared to $43000$ single photon events
in one day which should be sufficient for an experimental observation.
The coincidence detection of rare two-photon events, where both photons are emitted within a
small opening angle has been successfully demonstrated in the photon splitting process
\cite{Budker} ten years ago.
{The experimental sensitivity might be increased by a simultaneous detection of the scattered
electrons (like in photon tagging)}.
The electron beam should be a dilute beam tuned to one interaction per laser pulse.
}

In the weak-field regime the rate of the two-photon Compton process is suppressed by a factor of $\alpha (k\cdot p/m^2)^2$ relative to the one-photon Compton process for
$k\cdot p\ll m^2$ \cite{Heitler:Physica1934}, as for our kinematics.
For the momenta considered here the estimated
suppression in the weak field regime is $3\times10^{-5}$.
To discuss the relevance
of the two-photon emission in strong laser fields
we define the two- to one-photon ratio as
$\mathcal R = (\frac{dW^{(2)}}{d\Omega_1})/(\frac{dW^{(1)}}{d\Omega_1})$.
For $\tau=10$ we obtain a value of $\mathcal R = 10^{-2}$ at $a_0 = 1$ which is about two orders of magnitude larger than the perturbative estimate.
For lower values of $a_0< 0.1$ the suppression of the two-photon probability rapidly
approaches a constant value of $\mathcal R = 10^{-4}$ as anticipated in \cite{Heitler:Physica1934}.
Considering the different contributions we find that the ratio for the
on-shell process $\mathcal R_{\rm on} = 0.01  a_0^2$ for $a_0< 1$; the on-shell ratio is independent of $a_0$ for $a_0 < 0.1$, and above $a_0>0.1$ the value increases and
reaches $\mathcal R_{\rm off} = 10^{-3}$ at $a_0=1$.

Our approach furthermore opens the avenue towards a detailed study of the two-photon polarization which is considered
as a signature of the Unruh effect in \cite{Chen:Tajima:PRL1999,Schutzhold:PRL2008}.
The two-photon emission as a QED process in itself is also interesting with respect to the
quantum radiation reaction \cite{diPiazza:PRL2010}
as multiple incoherent one-photon Compton scatterings.
The sequential Compton scattering appears thereby
as a factorization of the resonant on-shell part of the scattering matrix element
on the amplitude level and is complemented by the possibility of coherent emission
due to the off-shell part.
Our numerical calculations show that up to 30 \% of the total photons are due to the off-shell
process and therefore beyond a description based solely on real intermediate photons.

In summary we provide the first complete evaluation of the differential probability
of two-photon emission by an electron in a
short intense laser pulse.
The on-shell part of the matrix element factorizes into subsequent one-photon
Compton processes and gives naturally a finite contribution to the differential probability 
due to the temporal pulse structure.
Our result allows for the first time an unambiguous comparison of the probability
of the two-photon process in relation to the one-photon process.
We find an increased two-photon yield by two orders of magnitude as compared
to the perturbative estimate of two-photon Compton scattering even for moderately strong
laser fields $a_0 \sim 1$ which are available presently at various laser facilities.

The authors gratefully acknowledge discussions with T.~E.~Cowan, R.~Sauerbrey, R.~Sch{\"u}tzhold and T.~St{\"o}hlker.

\end{document}